\documentclass[twocolumn]{article}
\usepackage[a4paper, total={180mm, 243mm}]{geometry}
\usepackage[utf8]{inputenc}
\usepackage{amssymb}
\usepackage{bm}
\usepackage{soul}
\usepackage{lipsum}
\usepackage{graphicx}
\usepackage{hyperref}
\usepackage[
    backend=biber,
    style=nature,
    url=false,
    isbn=false,
    eprint=false,
    date=year,
    maxbibnames=99
]{biblatex}
\AtEveryBibitem{%
  \clearfield{note}%
}
\usepackage[T1]{fontenc}
\usepackage{siunitx}
\usepackage{upgreek}
\usepackage[version=4]{mhchem} 
\usepackage{titling}
\usepackage{circledsteps}   
\usepackage{subcaption}
\usepackage{mathtools}
\usepackage{bm}
\usepackage[inkscapeformat=pdf]{svg}

\usepackage[version=4]{mhchem}  
\usepackage[switch]{lineno} 
\modulolinenumbers[5]

\DeclareUnicodeCharacter{2009}{\,} 

\DeclareCaptionLabelSeparator{custom}{\textbar}
\DeclareCaptionFormat{custom}{\textsf{\textbf{#1 #2} \small #3}}
\newcommand\authormark[1]{\textsuperscript{#1}}

\graphicspath{{./figures}} 

\makeatletter
\newcommand{\showfontsize}{\f@size{} pt}
\makeatother

\title{Spectral dynamics in broadband frequency combs \\ with overlapping harmonics}
\author{Weichen Fan\authormark{1}, Furkan Ayhan\authormark{2}, Thibault Wildi\authormark{1}, Mikhail Volkov\authormark{3}, \\ 
Ali Seer\authormark{3},  
Markus Ludwig\authormark{1,$\dag$}, 
Thibault Voumard\authormark{1,$\ddag$},
Andreas Brodschelm\authormark{3}, \\
Victor Brasch\authormark{4}, Guillermo L. Villanueva\authormark{2}, Tobias Herr\authormark{1,5,*}}

\date{%
    \small $^1$Deutsches Elektronen-Synchrotron DESY, Notkestr. 85, 22607 Hamburg, Germany \\
    \small $^2$\'Ecole Polytechnique F\'ed\'erale de Lausanne (EPFL), 1015 Lausanne, Switzerland \\
    \small $^3$ TOPTICA Photonics, 82166 Graefelfing / Munich, Germany\\
    \small $^4$ Q.ANT GmbH, Handwerkstraße 29, 70565 Stuttgart, Germany\\
    \small $^5$Physics Department, Universität Hamburg UHH, Luruper Chaussee 149, 22607 Hamburg, Germany\\
    $^{\dag}$ Presently: Université du Luxembourg, 162a, Avenue de la Faïencerie, L-1511 Luxembourg, Luxembourg\\
    $^{\ddag}$ Presently: Centre Suisse d'\'Electronique et de Microtechnique CSEM, Rue Jaquet-Droz 1, 2002 Neuchâtel, Switzerland\\
    $^*$tobias.herr@desy.de \\
}

\begin{document}

\maketitle

\textbf{
Optical frequency combs and their spectra of evenly spaced discrete laser lines are essential to modern time and frequency metrology. Recent advances in integrated photonic waveguides enable efficient nonlinear broadening of an initially narrowband frequency comb to multi-octave bandwidth. Here, we study the nonlinear dynamics in the generation of such ultra-broadband spectra where different harmonics of the comb can overlap. We show that a set of interleaved combs with different offset frequencies extending across the entire spectrum can emerge, which transform into a single evenly spaced ultra-broadband frequency comb when the initial comb is offset-free. 
}

\subsection*{Introduction}
Optical frequency combs and their discrete spectra enable phase-coherent links across the electromagnetic spectrum and underpin some of the most advanced measurements in physics \cite{fortier201920YearsDevelopments, diddams2020OpticalFrequencyCombs}. Usually, they are derived from femtosecond pulsed lasers and their frequency components are described by $\nu_m = m f_\mathrm{rep} + f_\mathrm{ceo}$, where $f_\mathrm{rep}$ and $f_\mathrm{ceo}$ are the laser's pulse repetition rate and carrier-envelope-offset frequency, and $m\in{\mathbb{N}_{0}}$ is the comb line index. While the initial comb spectra are often limited in span by the laser gain medium, nonlinear spectral broadening through self-phase modulation (SPM) in optical fibers \cite{dudley2006SupercontinuumGenerationPhotonic} has enabled octave spanning spectra \cite{ranka2000VisibleContinuumGeneration}, which are now routinely used to implement self-referencing i.e., detection of $f_\mathrm{ceo}$ as a beating between harmonics of the comb~\cite{telle1999CarrierenvelopeOffsetPhase, holzwarth2000OpticalFrequencySynthesizer, jones2000CarrierEnvelopePhaseControl, udem2002OpticalFrequencyMetrology}. 
Complementing silica fibers and specialty fibers~\cite{sylvestre2021RecentAdvancesSupercontinuuma}, as powerful nonlinear platforms waveguides have enabled access not only to SPM, but additional second order nonlinear processes such as sum and difference frequency generation (SFG/DFG) \cite{langrock2007GenerationOctavespanningSpectra}.
Especially in nanophotonic waveguides~\cite{bres2023SupercontinuumIntegratedPhotonics}, effects beyond SPM have enabled ultra-broadband spectra and power efficient implementation of $f_\mathrm{ceo}$ beatnote detection 
\cite{hickstein2017UltrabroadbandSupercontinuumGeneration, carlson2017SelfreferencedFrequencyCombs, okawachi2018CarrierEnvelopeOffset, hickstein2019SelforganizedNonlinearGratings, yu2019CoherentTwooctavespanningSupercontinuum,obrzud2019VisibleBluetored10, okawachi2020ChipbasedSelfreferencingUsing, lu2020UltravioletMidinfraredSupercontinuum,obrzud2021StableCompactRFtooptical, lesko2021SixoctaveOpticalFrequency, wu2024VisibletoultravioletFrequencyComb, ludwig2024UltravioletAstronomicalSpectrograph, fan2024SupercontinuaIntegratedGallium}, holding potential for optical spectroscopy and efficient phase coherent links from infrared to ultraviolet wavelengths.
Generally, in multi-octave spectra, the broadened fundamental comb can overlap with its harmonics that can result from e.g., concurrent SFG/DFG. This implies that, where different harmonics overlap, combs with different offset frequencies are interleaved and the equidistance of the frequency comb modes is broken. 

\begin{figure}[ht!]
    \centering
    \includegraphics[width=0.45\textwidth]{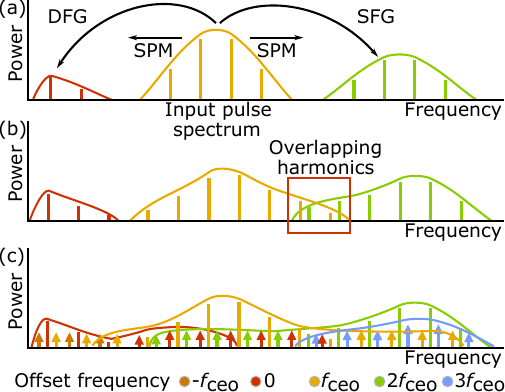}
    \caption{\textbf{Offset dynamics in broadband frequency combs with overlapping harmonics. (a)} Self-phase modulation (SPM), as well as sum- and difference frequency generation (SFG/DFG) result in spectral broadening and harmonic generation. \textbf{(b)} As the spectrum evolves, harmonics may overlap. \textbf{(c)} Harmonics with different offsets enter into additional nonlinear mixing processes, resulting in nonlinear gain for combs at new offset frequencies throughout the spectrum.}
    \label{fig_concept}
\end{figure}

Here, we study the nonlinear spectral dynamics in the generation of broadband comb spectra where an overlap between different harmonics occurs. Strikingly, we find that the interleaved combs do not remain confined to the narrow spectral band where harmonics initially overlap, but that they can extend across the entire spectrum. This can give rise to an entire set of multiple broadband interleaved combs. In the special case, where the initial comb is arranged to be offset-free ($f_\mathrm{ceo}=0$), a single frequency comb can be generated with evenly spaced comb lines spanning multiple octaves.
These findings are of immediate relevance to comb-based techniques that seek to leverage multi-octave spectra as those accessible in photonic-integrated waveguides.

\subsection*{Conceptual description}
Figure~\ref{fig_concept} illustrates the evolution of a broadband spectrum during propagation in a waveguide with second and third order nonlinearity; similar reasoning will also apply to a purely third order nonlinear waveguide with third-harmonic (triple-sum-frequency) generation: SPM broadens the input spectrum and the second order nonlinearity creates a SFG/DFG spectrum. If the broadening is sufficiently strong, the input comb with offset frequency $f_\mathrm{ceo}$, the SFG spectrum with offset frequency $2f_\mathrm{ceo}$, and the DFG spectrum with zero offset may start overlapping. In the overlapping spectral intervals this will result in interleaved combs with the same repetition rate $f_\mathrm{rep}$ but with different offsets.
As in this case the nonlinear gain window of different nonlinear processes overlap, there will be additional nonlinear mixing dynamics between combs that can in principle extend the interleaved combs across the entire spectrum, similar to related dynamics observed in nonlinear microresonators \cite{herr2012UniversalFormationDynamicsa}.
This could even result in the generation of additional interleaved combs with offsets corresponding to integer multiples of $f_\mathrm{ceo}$.
An experimental signature indicating the presence of broadband interleaved combs would be the detection of the offset frequency beatnote $f_\mathrm{ceo}$ in narrow spectral intervals across the entire spectrum, particularly outside the narrow spectral interval where the harmonics initially overlap. If multiple interleaved combs are present at the same spectral position, additional beatnotes between combs with offsets of integer multiples of $f_\mathrm{ceo}$ could be observed.

\subsection*{Numeric simulation}
To understand the nonlinear spectral dynamics in realistic waveguides, we perform numerical simulations \cite{voumard2023SimulatingSupercontinuaMixed} based on the approach introduced in \cite{conforti2013InteractionOpticalFields}. Specifically, we simulate a lithium niobate waveguide with a cross-section of 1000~nm$\times$800~nm and sidewall angle of \SI{75}{\degree}. Its group-velocity dispersion is derived through finite element methods using extraordinary refractive index of \ce{LiNbO3} from~\cite{zelmon1997InfraredCorrectedSellmeier} and refractive index of \ce{SiO2} from \cite{malitson1965InterspecimenComparisonRefractive}. The second-order nonlinearity $d_{33}$ is set as 31~pm/V and the nonlinear refractive index $n_2$ is set as $2.65\times10^{-19}$~m$^2$W$^{-1}$ \cite{phillips2011SupercontinuumGenerationQuasiphasematched}. The waveguide length is 5~mm of which the last third is periodically poled with a chirped poling period $\Lambda(z)$ ranging from \SI{1}{\um} to \SI{15}{\um} and 50\% duty cycle, to achieve broadband quasi-phase matching (QPM).
To spectrally resolve comb lines which belong to interleaved combs with different offsets, we use a periodic simulation time window that includes 32 sech$^2$-shaped input pulses with 120~pJ pulse energy, carrier frequency of 200~THz, 50~fs duration, and a periodicity of $f_\mathrm{rep}^{-1}=$ 375~ps, sufficiently long to avoid temporal overlap between consecutive pulses during propagation (To reduce computational effort it is also possible to simulate the spectra for each input pulse with different carrier-envelope offset phase separately). 
We define the input pulse train with $f_\mathrm{ceo}=f_\mathrm{rep}/4$, so that interleaved combs with offsets $0$, $f_\mathrm{ceo}$, $2f_\mathrm{ceo}$ and $3f_\mathrm{ceo}$ can be distinguished. Overall, the spectrum may contain frequencies $\nu_{m,n}=mf_\mathrm{rep}+nf_\mathrm{ceo}$, where we introduce the harmonic index $n\in{\mathbb{N}_{0}}$ (we have $n=1$ for the input pulse).

\begin{figure}[t!]
    \centering
    \includegraphics[width=0.4\textwidth]{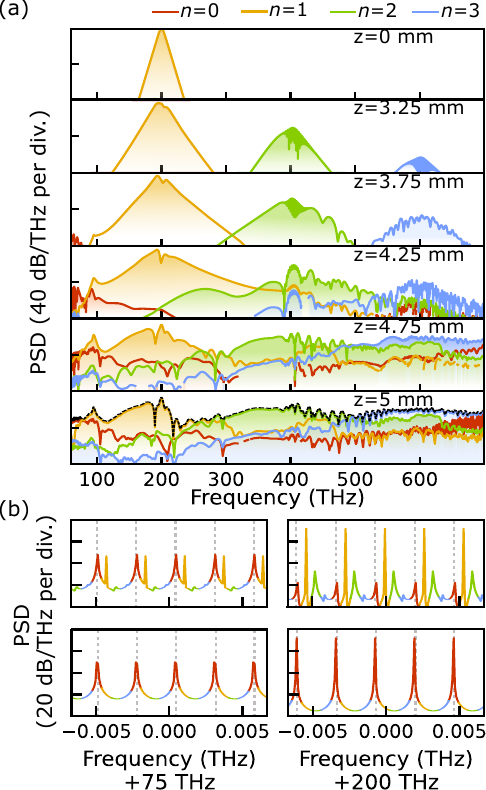}
    \caption{\textbf{Numerical simulation results.} \textbf{(a)} Spectral envelope evolution of comb spectra with different harmonic index $n$ along propagation z. Dashed black line indicates the output spectrum. PSD: power spectral density. \textbf{(b)} Magnified view of the output spectrum around 75~THz and 200~THz in the condition of: upper panels: $f_\mathrm{ceo}=f_\mathrm{rep}/4$, and lower panels: $f_\mathrm{ceo}=0$. Dashed grey lines mark the offset-free frequency position.
    }
    \label{fig_simu}
\end{figure}

From the simulated nonlinear dynamics, we extract the spectral envelope of each harmonic $n$ along the spatial propagation coordinate $z$ as shown in in Fig.~\ref{fig_simu}a. The input spectrum ($n=1$) broadens and initially well separated second ($n=2$) and third ($n=3$) harmonic spectra are generated, which then start to overlap. 
During continued propagation, including in the periodically poled portion of the waveguides, the harmonics spread throughout the entire spectrum, forming effectively a set of multiple interleaved broadband combs with different $n$.
The upper panels of Fig.~\ref{fig_simu}b show the spectra in a narrow spectral window around 75~THz and 200~THz, where the resolved interleaved combs are visible. In contrast, if the pulse train is defined to have $f_\mathrm{ceo}=0$, interleaved combs are absent as can be seen in the lower panels of Fig.~\ref{fig_simu}b showing the same spectral windows.

\subsection*{Experiments}
\begin{figure}[ht!]
    \centering
    \includegraphics[width=0.45\textwidth]{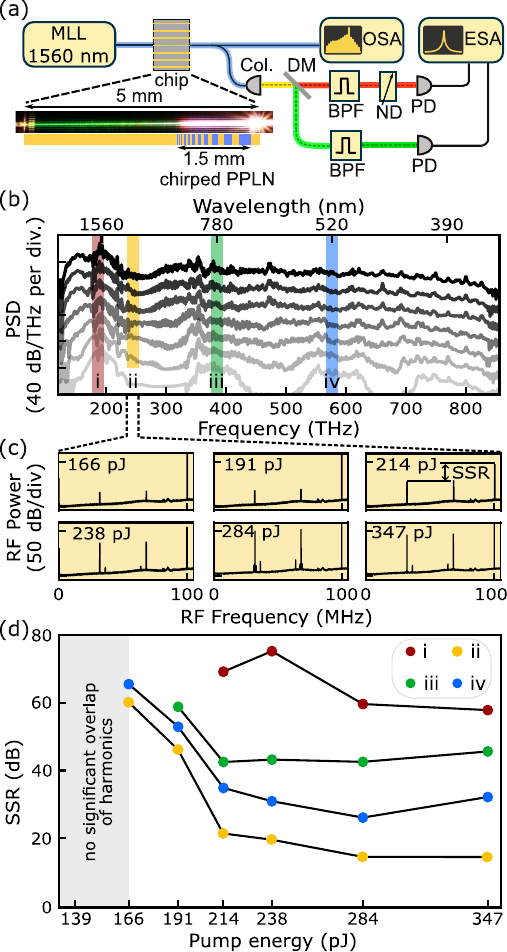}
    \caption{\textbf{Experiments with non-zero offset. (a)} Schematic setup and a photograph of the chip during operation. PPLN: periodically poled lithium niobate; OSA: optical spectrum analyzer; ESA: electrical spectrum analyzer; Col.: collimator; DM: dichroic mirror; BPF: bandpass filter; ND: neutral density filter; PD: photodetector. \textbf{(b)} Optical spectra with increasing pump energy (faded to dark: 139~pJ, 166~pJ, 191~pJ, 214~pJ, 238~pJ, 284~pJ, and 347~pJ). The spectra are shifted by 25~dB for visibility. Colored vertical bars mark frequency positions used to measure the RF beatnote, with (i) 1610$\pm$6~nm, (ii) 1300$\pm$15~nm, (iii) 780$\pm$5~nm, and (iv) 520$\pm$18~nm. 
    \textbf{(c)} The RF beatnote at around 1300~nm with increasing pump energy with resolution bandwidth (RBW) of 10~kHz. \textbf{(d)} Scaling of the sideband suppression ratio (SSR) with the pulse energy. Data points represent the ratio between $P_\mathrm{rep}$ and $P_\mathrm{ceo}$ at different frequencies marked in \textbf{(b)}.
    }
    \label{fig_experiment}
\end{figure}

In the experiments, we use a 5~mm long x-cut LNOI waveguide with \ce{SiO2} cladding that is pumped in the fundamental TE mode (polarization parallel to the optical axis of \ce{LiNbO3}). The waveguide has a cross-section of 2300~nm$\times$800~nm, and the last 1.5~mm (excluding the output coupling taper) of the waveguide is periodically poled with a chirped poling period as in the simulation. Detailed waveguide fabrication methods can be found in previous work \cite{ayhan2025FabricationPeriodicallyPoled}. The experimental setup is shown in Fig.~\ref{fig_experiment}a, where the pump laser is a sub-100~fs mode-locked laser with a center wavelength of 1560~nm and a repetition rate of 100~MHz.  The light is coupled into and out from the waveguides via lensed fibers, and all fiber components are polarization maintaining and single-mode at 1560~nm. The coupling efficiency is approximately 23\% per facet and the on-chip pulse energy is varied from 139~pJ to 347~pJ. Optical spectrum analyzers (OSAs) are used to record the broadband output spectra. For RF beatnote measurements, the light is collimated to free-space via an off-axis parabolic mirror, and a long-pass dichroic mirror (DM) splits the light with transmission wavelength starting from 950~nm to avoid potential ambiguity. Then the light is sent through bandpass filters (BPFs) before reaching the photodetectors (PDs). A neutral density filter (ND) is placed in the long wavelength optical path to avoid saturating the PD. The generated radio-frequency (RF) signals are measured by an electrical spectrum analyzer (ESA).

Figure~\ref{fig_experiment}b shows the output spectra observed for different pump pulse energies, similar to the spectral evolution with increasing propagation distance in the simulation. For the lowest pulse energy, the harmonics are well separated, but with increasing pulse energy merge into a gap-free spectrum spanning from mid-infrared to ultra-violet wavelengths. The RF signals are measured around the fundamental pump wavelength, its second harmonic and third harmonic, as well as around 1300~nm, as indicated by the color bands in Fig.~\ref{fig_experiment}b.

For the lowest pulse energy of 139~pJ, where harmonics are well separated, no $f_\mathrm{ceo}$ signal is observed in any wavelength band, despite the RF power of $f_\mathrm{rep}$ in the bands within the second and third harmonic regions exceeding the PD noise floor by more than 60~dB (see Fig.~\ref{fig_all_beatnote} in the Supplementary Information, SI). This changes with increasing pulse energy, as shown in Fig.~\ref{fig_experiment}c for the 1300$\pm$15~nm band and in the SI (Fig.~\ref{fig_all_beatnote}) for other wavelength bands: As the harmonic gaps close, an $f_\mathrm{ceo}$ signal emerges at 166~pJ pulse energy, marked by two RF peaks between 0 and $f_\mathrm{rep}=100$~MHz, indicating the formation of interleaved combs. For higher pulse energy the strength of the $f_\mathrm{ceo}$ signal increases and two additional offset beatnotes appear for pulse energy above 238~pJ, indicating the simultaneous presence of at least three interleaved combs at 1300~nm. 

The power level of the repetition rate and the offset frequency beatnotes from the interleaved combs may be used to estimate the relative power levels of the interleaved comb lines. To simplify this estimation, we assume that only two interleaved combs $n=i,j$ are present (i.e. any additional interleaved combs are of much less power) and that the spectral envelope and phase of the lines within each of the interleaved combs does not vary significantly across the bandwidth of the filter: Denoting with $A_{m,n}$ the (complex) field strength of the comb line with frequency $\nu_{m,n}$, the RF power levels in the repetition rate beatnote and the offset frequency beatnotes are:

\begin{equation}
\label{eq:Prep}
\begin{split}
    P_\mathrm{rep} & = \eta (|A_{m,i}A_{m-1,i}^*|^2+|A_{m,j}A_{m-1,j}^*|^2) \\
    & \approx \eta (P_{m,i}^2+P_{m,j}^2),
\end{split}
\end{equation}

\begin{equation}
\label{eq:Pceo}
\begin{split}
    P_\mathrm{ceo} & = \eta (|A_{m,i}A_{m,j}^*|^2+|A_{m-1,i}A_{m-1,j}^*|^2) \\
    & \approx 2\eta P_{m,i}P_{m,j},
\end{split}
\end{equation}
where lines with $m$ and $m-1$ are assumed to pass through the spectral filter, and $\eta$ is a constant that depends on the filter bandwidth and the detection efficiency. Assuming the interleaved comb with $n=i$ is significantly more powerful than the one with $n=j$, we define the sideband suppression ratio (SSR) of the dominant comb as the power ratio between the two combs:
\begin{equation}
    \label{eq3}
    \mathrm{SSR} \coloneqq \frac{P_{m,i}}{P_{m,j}} \approx \frac{2P_\mathrm{rep}}{P_\mathrm{ceo}} 
\end{equation} 
where in the experiment the $\mathrm{SSR}$ is limited by the RF noise floor which defines a minimum value of $P_\mathrm{ceo}$. Figure~\ref{fig_experiment}d shows the SSR obtained through this estimation for the different wavelength bands. The SSR drastically drops for very broadband spectra and may be as low as 15~dB at 1300~nm. 
Such low SSR (low power contrast) may become problematic for some applications like dual-comb spectroscopy \cite{coddington2016DualcombSpectroscopy,picque2019FrequencyCombSpectroscopy}, spectroscopy based on individual comb line detection \cite{diddams2007MolecularFingerprintingResolved} or astronomical spectrograph calibration \cite{herr2019AstrocombsRecentAdvances,jovanovic20232023AstrophotonicsRoadmapa}. However, as the coherence of the comb lines is preserved (narrow beatnotes are detected), phase coherent links across the comb spectrum may still be implemented if the interleaved comb line spectrum is accounted for.

\begin{figure}[ht!]
    \centering
    \includegraphics[width=0.45\textwidth]{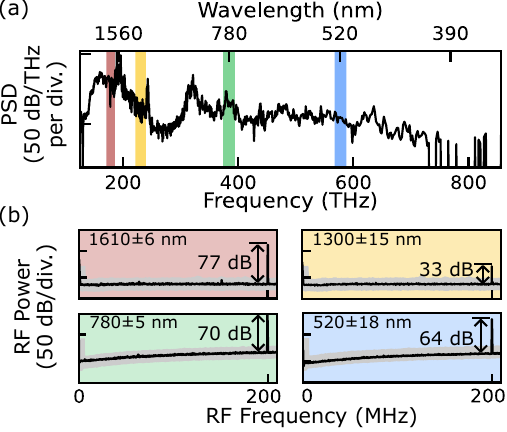}
    \caption{\textbf{Experimental results with offset-free pump source. (a)} Generated optical spectrum. \textbf{(b)} RF beatnote measured at optical frequencies same as in the non-offset-free pump experiments. Grey backgrounds mark the noise floor of PDs. RBW=1~kHz. 
    }
    \label{fig_offset_free}
\end{figure}

In agreement with the simulations, a strictly equidistant frequency comb can be generated when the pump source is arranged with $f_\mathrm{ceo}=0$. To demonstrate this experimentally, we pump the waveguides with an amplified offset-free frequency comb source (TOPTICA DFC CORE 200 +) based on a DFG comb \cite{krauss2011AllpassivePhaseLocking}, providing 472~pJ pulse energy (on-chip) at a center wavelength of 1560~nm with 59~fs pulse duration at 200~MHz repetition rate and aligned in TE polarization. The waveguide has the same parameters as the one in aforementioned experiments, except the cross-section is 1800~nm$\times$800~nm which results in the broadest spectrum at this pump configuration (Fig.~\ref{fig_offset_free}a). Indeed, as expected, the RF measurements do not show any detectable offset signal in all filter bands (Fig.~\ref{fig_offset_free}b), in agreement with the absence of interleaved combs.

\subsection*{Conclusion}
Highly efficient nonlinear waveguides enable ultra-broadband frequency combs. Overlap between harmonics can occur in such spectra, leading to a set of interleaved combs with different offset frequencies and these interleaved combs can extend across the entire spectrum. While this effectively results in an non-equidistant spectrum of lines, the frequency components are still well-defined and we do not observe a loss of coherence between the comb lines.  Thus, if the interleaved combs are taken into account, these spectra are still suitable for phase-coherent links across the spectrum.
For comb-based techniques that rely on the regularity of the comb spectrum and absence of interleaved lines, such as dual-comb spectroscopy, spectroscopy based on individual comb line detection, or astronomical spectrograph calibration, overlap between harmonics should be avoided. Alternatively, when the waveguide is pumped with an offset-free source~\cite{krauss2011AllpassivePhaseLocking, okubo2018OffsetfreeOpticalFrequency}, a single self-consistent gap-free ultra-broadband frequency comb with evenly spaced comb lines emerges, creating new opportunities for ultra-broadband optical precision across multiple-octaves.

\subsection*{Acknowledgments}
This project has received funding from the European Research Council (ERC, grant No 853564), the European Innovation Council (EIC, grant No 101046920), the Swiss National Science Foundation (Sinergia, grant No 00020\_182598) and through the Helmholtz Young Investigators Group VH-NG-1404. The fabrication of the lithium niobate waveguides was carried out at the Center for MicroNanoTechnology (CMi) at EFPL, and simulations were supported through the Maxwell computational resources operated at DESY.

\printbibliography

\clearpage

\onecolumn

\title{Supplementary Information to: Spectral dynamics \\
in broadband frequency combs with overlapping harmonics}
\author{}
\date{}
\maketitle

\setcounter{section}{0}
\setcounter{figure}{0}
\setcounter{equation}{0}
\renewcommand{\theequation}{S\arabic{equation}}
\renewcommand{\thefigure}{S\arabic{figure}}

\begin{figure}[ht!]
    \centering
    \includegraphics[width=0.75\textwidth]{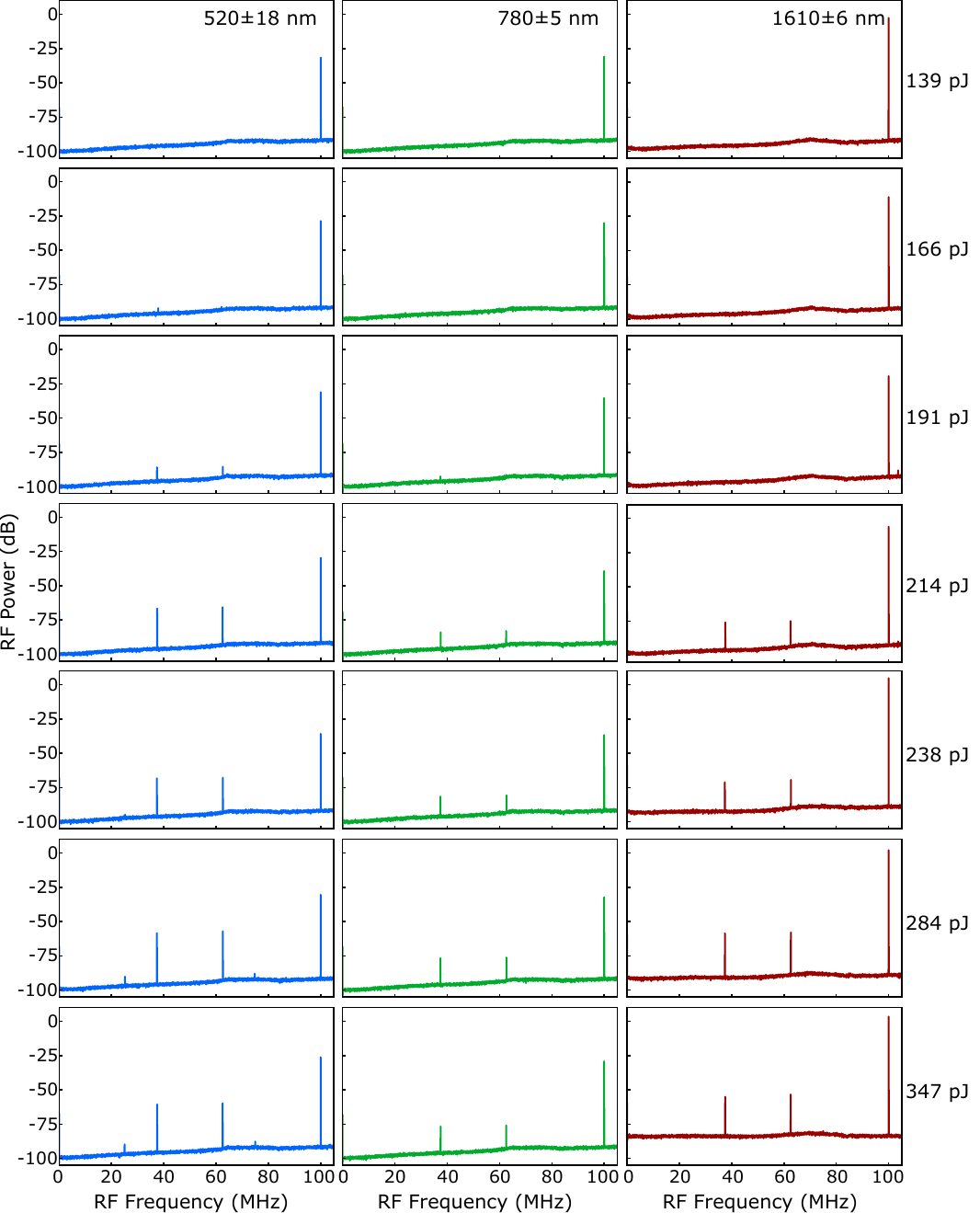}
    \caption{Evolution of all RF spectra obtained after photodetection of the bandpass filtered spectra around fundamental, second harmonic, and third harmonic with increasing pump energy. Resolution bandwidth 10~kHz. Video bandwidth 30~Hz.}
    \label{fig_all_beatnote}
\end{figure}

\end{document}